\documentclass[letter]{aa} 
\usepackage{graphicx}
\usepackage{natbib}
\usepackage{txfonts}
\begin{document}
   \title{Oscillation mode linewidths of main-sequence and subgiant stars observed by {\it Kepler}}

   \author{T.~Appourchaux
          \inst{1}
          \and
          O.~Benomar\inst{2}
          \and
          M.~Gruberbauer\inst{3}
          \and
         W.~J.~Chaplin\inst{4}
          \and
          R.~A.~Garc\'\i a\inst{5}
          \and
          R.~Handberg\inst{6}
          \and
         G.~A.~Verner\inst{5}
          \and
          H.~M.~Antia\inst{7}
          \and
          T.~L.~Campante\inst{6,8}
          \and
          G.~R.~Davies\inst{5}
          \and
          S.~Deheuvels\inst{9}
          \and
          S.~Hekker\inst{10,4}
          \and
          R.~Howe\inst{4}
          \and
          D.~Salabert\inst{11}
          \and
          T.~R.~Bedding\inst{2}
          \and
          T.~R.~White\inst{2}
          \and
          G.~Houdek\inst{12}
          \and
          V.~Silva Aguirre\inst{13}
          \and
          Y.~P.~Elsworth\inst{4}
          \and
          J.~Van Cleve\inst{14}
          \and
          B.~D.~Clarke\inst{14}
          \and
          J.~R.~Hall\inst{15}
          \and
          H.~Kjeldsen\inst{6}
          }

   \institute{Univ Paris-Sud, Institut d'Astrophysique Spatiale, UMR8617, CNRS, B\^atiment 121, 91405 Orsay Cedex, France
          \and
          Sydney Institute for Astronomy (SIfA), School of Physics, University of Sydney, New South Wales 2006, Australia
          \and
          Department of Astronomy \& Physics, Saint Mary's University, Halifax, NS B3H 3C3, Canada
          \and
          School of Physics and Astronomy, University of Birmingham, Edgbaston, Birmingham B15 2TT, United Kingdom
          \and
          Laboratoire AIM, CEA/DSM-CNRS-Universit\'e Paris Diderot, IRFU/SAp, Centre de Saclay, 91191 Gif-sur-Yvette Cedex, France
          \and
          Department of Physics and Astronomy, Aarhus University, Ny Munkegade 120, DK-8 Aarhus C, Denmark
          \and
          Tata Institute of Fundamental Research, Homi Bhabha Road, Mumbai 45, India
          \and
          Centro de Astrof\'isica, DFA-Faculdade de Ci\^encias, Universidade do Porto, Rua das Estrelas, 4150-762 Porto, Portugal
          \and
         Department of Astronomy, Yale University, P.O. Box 208101, New Haven CT 06520-8101, USA
          \and
          Astronomical Institute "Anton Pannekoek", University of Amsterdam, Science Park 904, 1098 XH Amsterdam, The Netherlands%
          \and
          Universit\'e de Nice Sophia-Antipolis, CNRS UMR 6202, Observatoire de la C\^ote d'Azur, BP 4229, 06304 Nice Cedex 4, France
          \and
          Institute of Astronomy, University of Vienna, A-1180 Vienna, Austria
          \and
          Max Planck Institute for Astrophysics, Karl-Schwarzschild-Str. 1, 85748, Garching bei M\"{u}nchen, Germany
          \and
          SETI Institute/NASA Ames Research Center, Moffett Field, CA 94035
          \and
          Orbital Sciences Corporation/NASA Ames Research Center, Moffett Field, CA 94035
          }

   \date{Received 22 November 2011; accepted 14 December 2011}

 
  \abstract
  {Solar-like oscillations have been observed by {{\it Kepler}} and CoRoT in several solar-type stars.}
   {We study the variations of stellar  p-mode linewidth as a function of effective temperature.}
   {Time series of 9 months of {{\it Kepler}} data have been used.  The power spectra of 42 cool main-sequence stars and subgiants have been analysed using both Maximum Likelihood Estimators and Bayesian estimators, providing individual mode characteristics such as frequencies, linewidths and mode heights.}
   {Here we report on the mode linewidth at maximum power and at maximum mode height for these 42 stars as a function of effective temperature.}
   {We show that the mode linewidth at either maximum mode height or maximum amplitude follows a scaling relation with effective temperature, which is a combination of a power law plus a lower bound.  The typical power law index is about 13 for the linewidth derived from the maximum mode height, and about 16 for the linewidth derived from the maximum amplitude while the lower bound is about 0.3 $\mu$Hz and 0.7~$\mu$Hz, respectively.  We stress that this scaling relation is only valid for the cool main-sequence stars and subgiants, and does not have predictive power outside the temperature range of these stars.}
   \keywords{stars : oscillations, {{\it Kepler}}
               }

   \maketitle
%

\section{Introduction}
\label{sec:intro}

Stellar physics faces a revolution following the great wealth of asteroseismic data made available by space missions such as CoRoT \citep{Baglin2006} and {\it Kepler} \citep{Borucki2009}.   Long observations of solar-like pulsators corresponding to main sequence stars, subgiants and red giants have been performed during more than 6 months by CoRoT \citep[][and references therein]{Baudin2011,Baudin2011C}.  The {\it Kepler} mission now provides a larger sample of stars observed for longer durations \citep{Chaplin2011}.

The study of oscillation mode physics (mode height, linewidth and amplitude) provides information on excitation and damping mechanisms related to the physics of convection and of stellar atmospheres \citep{Samadi2009}. 
\citet{GH99} theoretically derived stellar mode linewidths as a function of stellar mass and age.  They found that stellar mode linewidths would present a depression or {\it plateau} close to the maximum of mode height.  This depression was first detected in the solar p-mode linewidths by \citet{CFJR95}.  Such a depression is due to a resonance between the thermal adjustment time of the superadiabatic boundary layer and the mode frequency \citep{B92a}.   
The frequency location of the maximum of mode height is in turn 
related to the Mach number (${\cal M}_a$), the ratio of convective velocity to the sound speed \citep{Belkacem2011}.  The convective flux giving the maximum mode amplitude is also related to ${\cal M}_a$ to the power of 3 \citep[e.g.][]{Belkacem2011,GH99}.  It is therefore interesting to study how the mode linewidth is related to the frequency of maximum amplitude / mode height for several different stars.

Statistical studies over a large number of stars have been performed in order to  validate the scaling relation derived for the amplitude of stellar oscillations by \citet{HK1995} and recently revised by \citet{KB2011}.    Scaling relations for mode linewidth have been proposed by \citet{Chaplin2009} and \citet{Baudin2011} based upon the stellar effective temperature.  


 \citet{Chaplin2009} proposed a scaling relation with linewidth proportional to $T_{\rm eff}^4$ based upon several ground-based observations.  Using CoRoT observations, \citet{Baudin2011C} measured linewidths for a sample of solar-like pulsators and provided a scaling relation proportional to $T_{\rm eff}^{16}$.

With the ability to perform longer observations of stars with {\it Kepler}, the measurement of mode linewidth becomes easier and more reliable.  
In this paper, based upon {\it Kepler} observations and using a larger stellar sample than of \citet{Baudin2011}, we derive a new relation between mode linewidth and $T_{\rm eff}$ .


\section{Data analysis}
\label{sec:method}

\subsection{Time series and power spectra}
{\it Kepler} observations are obtained in two different operating modes: long cadence (LC) and short cadence (SC) \citep{Gilliland2010,Jenkins2010}.  This work is based on SC data. For the brightest stars (down to {\it Kepler} magnitude, $Kp \approx12$), SC observations can be obtained for a limited number of stars  (up to 512 at any given time)  with a faster sampling rate of 58.84876~s (Nyquist frequency of $\sim$ 8.5~mHz), allowing for more precise transit timing.   The time series were corrected for outliers, occasional jumps and drifts \citep[see][]{RAG2011}, and the levels between the quarters were normalized.  Finally, the resulting light curves have been high-pass filtered using a triangular smoothing of width 1 day, to minimize the effects of the long period instrumental drifts.   The power spectra were produced from a single source using the Lomb-Scargle periodogram \citep{Scargle82}, properly calibrated to comply with
Parseval's theorem \citep[see][]{Appourchaux2011}.

\begin{figure}[h]
   \centering
   \includegraphics[angle=0,width=8.25cm]{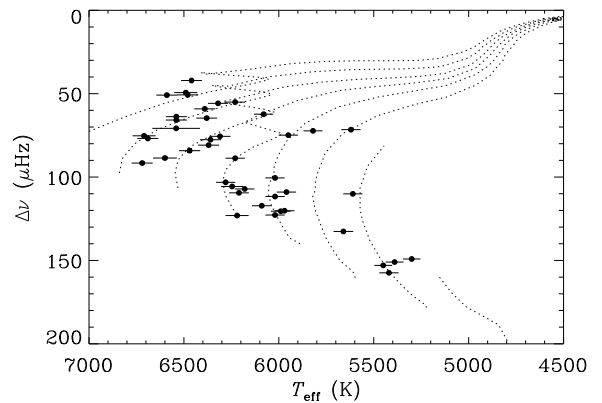}
      \caption{Large separation as a function of effective temperature of the stars used in this study.    The error bars on the large separations are within the thickness of the symbol.  The evolutionary tracks for stars of mass 0.8 M$_{\odot}$(most right) to  1.5 M$_{\odot}$ (most left) (by step of 0.1 M$_{\odot}$) are shown as dotted lines.  The tracks are derived from \citet{Marigo2008}.
              }
         \label{fig1}
   \end{figure}

{\it Kepler} observations are divided into three-month-long {\it Quarters} (Q).  A subset of 42  cool main-sequence stars and subgiants observed during quarters Q5, Q6 and Q7 (March  22, 2010 to December 22, 2010) were chosen for having oscillation modes with high signal-to-noise ratios ranging from 1.8 to 50 in the power spectrum.  The frequency resolution is about 0.04~$\mu$Hz.  
Figure~\ref{fig1} shows the measured large frequency separation of these 42 stars as a function of their effective temperature provided by \citet{Pins2011}.  The large separation is derived from individual mode frequencies at $\nu_{\rm max}$ from the {\it All} data set (See Table~\ref{table:0}.).  We took care to analyse solar-type stars without avoided crossings, since these may reduce the observed linewidths.  The avoided crossings were detected by visual inspection of the echelle diagram; examples of such avoided crossings can be found in  \citet{Deheuvels2010,Metcalfe2010, Mathur2011, Campante2011,Bedding2011}.

\subsection{Mode parameter extraction}
The mode parameter extraction was performed by 11 fitters.  The list of fitted modes were compared for completeness and 5 fitters were selected for finalising the parameters:  two fitters (IAS, BIR), who applied maximum likelihood estimators (MLE), and three Bayesian fitters (SYD, MAR and AAU).

The power spectra were modelled over a frequency range covering typically about 15 to 20 large separations (=$\Delta\nu$).  For each radial order, the model parameters were mode frequencies (one each for $l$=0,1,2), a single mode height (with an assumed ratio of $H_1/H_0=1.5$, $H_2/H_0$=0.5) and a single mode linewidth.  In the case of AAU only, the $l=0$ linewidths were fitted and the linewidths of the other degrees were interpolated in between two $l=0$ mode linewidths.  The relative heights $H_{(l,m)}$ of the split components of the modes depend on the stellar inclination angle as given by \citet{Gizon2003}.  For 
each star, the rotational splitting and stellar inclination angle were chosen to be common across all the 
modes.
The mode profile was assumed to be Lorentzian.  The background was modelled using a multi-component Harvey model \citep{Harvey85} with two parameters  and a white noise component.  We used a single Harvey component for all stars, and a double component for 11 stars (BIR's stars).  In total the number of free parameters for 15 orders was at least $5 \times 15+2=77$. 

The two models described above were used for fitting the parameters of the stars using MLE.  All 42 stars were fitted by IAS, 16 of which were fitted by IAS alone.  Eleven stars were fitted by BIR.  The fit was done without and with rotational splitting;  the significance of splitting and angle was then tested using the likelihood ratio test, by applying the H$_0$ hypothesis with a cutoff for a $\chi^2$ with 2 d.o.f of $\Delta{\log}({\rm likelihood})$=9.2 (4.6) or a probability of 10$^{-4}$ (10$^{-2}$) for IAS, and for BIR, respectively.   Formal uncertainties on each parameter were derived from the inverse of the Hessian matrix \citep[for more details on MLE, significance and formal errors, see][]{Appourchaux2011}. 

Fifteen stars with large mode linewidths were fitted with a Bayesian approach using different sampling methods. SYD and AAU employed MCMC \citep{Benomar2009,Handberg2011}, while MAR used nested sampling via the code MultiNest \citep{Feroz2009}. For the nested sampling approach, the large number of parameters forced us to use MultiNest's constant efficiency, mono-modal mode. The priors on the central frequency and inclination angle were uniform. The prior on the splitting was either uniform from 0-10~$\mu$Hz (MAR) or a combination of a uniform prior over 0-2 $\mu$Hz and a decaying Gaussian (SYD, AAU). The priors on mode height were modified Jeffreys priors \citep[][]{Benomar2009,Gruberbauer2009}, and the priors on the linewidth were either uniform (MAR) or modified Jeffreys priors (SYD, AAU). 
The error bars were derived from the marginal posterior distribution of each parameter.  Each Bayesian fitter had 7 stars to fit: 4 stars + 3 common stars.  The latter are used for comparison of the Bayesian methods.  Priors on frequencies were set after visual inspection of the power spectrum. Modes of degree $l=2$ were assumed to be on the low-frequency side of the $l=0$ (\emph{i.e.}, the small spacing $d_{02}$ is assumed positive).  In order to avoid spurious results, one of the Bayesian fitters (SYD) also used a smoothness condition on the frequency for each degree.

The different data sets available are summarized in Table~\ref{table:0}.

   \begin{table}[t]
\caption{Data set of fitted stars.}             
\label{table:0}      
\centering                          
\begin{tabular}{c c c c c}        
\hline                 
\hline     
Dataset&Fitter &Method&\# of stars &Comment \\    
\hline  
\hline                           
I&IAS & MLE	& 16 & No common stars \\
II&BIR &MLE & 11 & No common stars \\
III&SYD &Bayes&7  & Common stars$\dagger$\\
IV&MAR & Bayes&7 & Common stars$\dagger$\\
V&AAU & Bayes &7 &Common stars$\dagger$\\ 
All&IAS & MLE & 42 & All stars included \\
\hline
\hline  
\end{tabular}
$\dagger${From these, 3 stars commonly fitted by SYD, MAR and AAU}
\end{table}

\subsection{Linewidths}

In a similar fashion to \citet{Baudin2011}, we derived the mean linewidth ($\Gamma_{\nu_{\rm max}}$) at maximum mode height {\it and} at maximum mode amplitude by taking the weighted average of 3 linewidths of 3 orders around the frequency of these maxima (See Tables~\ref{table_mode_height} and \ref{table_amplitude} as online materials).   The derivation of $\Gamma_{\nu_{\rm max}}$ is rather immune to systematic effects resulting from the 3-mode average because at these frequencies the observed linewidths exhibit a {\rm plateau}, as shown theoretically by \citet{GH99}  and as observed in the Sun by \citet{CFJR95}.

Individual mode linewidths can have systematic errors resulting from the incorrect estimation of several mode profile parameters.  In addition, an over- or underestimation of mode linewidths will provide an under- or overestimation of mode heights, respectively.   Estimates of such systematic errors can be derived using the procedure developed by \citet{TT2005b}, which consists in fitting one model profile, without using Monte-Carlo simulations.  

The main parameters producing systematic errors on mode linewidths are: the background noise $B$, the mode height ratio and the splitting.  

The major source of systematic errors on mode height and mode linewidth is the biased estimation of the background noise.  An estimate of the mode linewidth bias can be derived for a single mode using the analytical formulae provided by \citet{TTTA94}.  We can then derive the bias on mode linewidth as a function of the error on $\Delta B$ and the inverse signal-to-noise ratio ($\beta=B/H$) in the power spectrum:
\begin{equation}
\frac{\Delta \Gamma}{\Gamma}=k(\beta,\Gamma,\Delta\nu)\frac{\Delta B}{B}~,
\end{equation}
where $\Delta\nu$ is the window over which the fit is performed for that single mode.  Typically $k$ is negative and of order 1, i.e., under-estimation of the background by 10~\% will lead to an over-estimation of the linewidth by 10\%.  Another source of systematic errors is the assumption that the ratios of mode height be fixed to some given values.  There is indeed a variation of mode height ratios with effective temperature as shown by \citet{Ballot2011}.   The resulting underestimation of these ratios is typically not larger than 0.1, which corresponds roughly to an underestimation of the linewidths not larger than 3\%.  
A minor source of systematic errors comes from the rotational splitting.  In the case for which the splitting is not detected (typically when the splitting is not greater than 10~\% of the linewidth), the linewidth will be overestimated by about 6\% for $\Gamma~=10~\mu$Hz, and by about 3\% for $\Gamma~=~3~\mu$Hz.  When the splitting is larger, there is no correlation between the detected splitting and the linewidth \citep{TTTA94}.  All these values were either confirmed or inferred with the procedure suggested by \citet{TT2005b}.

Last but not least, an extrinsic systematic effect on the linewidth is related to stellar activity.  It was shown by \citet{WC2000}, that the solar linewidth may change by typically 20\% at the location of the dip.  We are aware that this can have an effect on the mean linewidth reported here.  For many stars, this effect cannot be assessed with such a short observation duration of 9 months.




\section{Discussion}
\begin{figure}
   \centering
   \includegraphics[angle=+90,width=8.25cm]{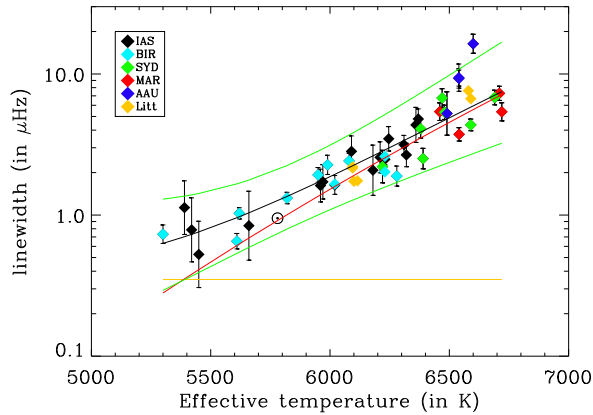}
      \caption{Average mode linewidth at maximum mode height (and their 3-$\sigma$ error bars) as a function of effective temperature  \citep[provided by][]{Pins2011}.  The error bars on the effective temperatures although not shown here are indeed included in the error analysis.  Average mode linewidth fitted by
      IAS (black), by BIR (cyan), by SYD (green), by MAR (red), by AAU (blue), from \citet{Baudin2011} (orange).  
      Fitted average linewidth (Black line).  3-$\sigma$ error bars on fitted average linewidth (Green lines).  Power law component of the fit (Red line).  Flat component at low  $T_{\rm eff}$ (Orange line).  The mean mode linewidth of the Sun is indicated at 5777 K.
              }
         \label{fig3}
   \end{figure}

Figure \ref{fig3} shows the linewidth measured at maximum mode height as a function of effective temperature.  
We note that \citet{Chaplin2009}
proposed a scaling relation, which provides a variation of the mode linewidth by a factor 2.7 between 6800~K and 5300 K; while 
\citet{Baudin2011C}
provides a factor of 53.9 for the same temperature change.  The measured ratio, here, is closer to 10. It is clear that neither dependence is adequate to explain our measurements.  The results of \citet{Chaplin2009} were based on predicted mode lifetimes from pulsation computations, and also on a small number of relatively short ground-based observations, potentially subject to large systematic errors.

We tested three forms of the $T_{\rm eff}$ relations, namely an exponential variation, a pure power law, and a power law with a flat component.  Without any physical basis for choosing between the different relations, we adopted the one with the lowest $\chi^2$, which was the third of these:
\begin{equation}
	\Gamma=\Gamma_0+\alpha \left(\frac{T_{\rm eff}}{5777}\right)^s~.
\label{Eq1}	
\end{equation}
The effective temperatures were derived from two re-calibrations of the photometry in the {\it Kepler} Input Catalog: one based on $griz$ use of the photometry \citep{Pins2011} and one based on application of the Infrared Flux Method using 2MASS $JHK$  \citep{Casa2010, Casa2006}.  The random errors on the fitted parameters were derived using Monte-Carlo simulations of the fit taking into account random errors on both the effective temperature and the linewidths.   

Tables~\ref{table:1} and \ref{table:2} give the results of the fitted parameters of the linewidth at $\nu_{\rm max}$ for the two different effective temperatures and the two different ways of measuring $\nu_{\rm max}$.  This latter can be derived either from the maximum of the mode amplitude, which is $\propto E$ (where $\sqrt{E}$ is the energy injected by convection) or from the maximum of mode height, which is $\propto E/\Gamma$.  We used five different sets of linewidth data to study the impact of the different method upon the fitted parameters:  all fitted linewidth (MLE and Bayesian), all fitted linewidth (excluding either BIR or IAS's), MLE (fitted by IAS and BIR only), MLE (fitted only by IAS). 

Here we note that the power law indices are rather close to the index given by \citet{Baudin2011C} (See Table \ref{table:1}).  The mode linewidth measured at the maximum mode height is systematically lower on average by about 10\% than that measured at the maximum amplitude.  This is because the frequency of maximum amplitude tends to be higher than the frequency of maximum mode height.


The different power law index between the two sources of effective temperature is mainly due to the fact that the range of temperature is smaller for \citet{Pins2011} compared to \citet{Casa2010}; the reduction is 75 K, mainly at the high temperatures.  The lower temperature range would increase $s$ by 1.0 and 1.5, which is roughly in agreement with Tables \ref{table:1} and \ref{table:2}, respectively.

We also studied the impact of having different fitters upon the derived parameters.  From Tables~\ref{table:1} and \ref{table:2}, we can see that the fitted parameters are the same within error bars when we combine the MLE fits with the Bayesian fits.  There is a much larger difference when we use the linewidth derived on {\it all} stars by IAS only (the only homogenous data set), thereby also including the stars for which the effective temperature is higher than 6400 K.  For that homogenous data set, the linewidths measured at high effective temperature are systematically higher than those measured by the Bayesian fitters by up to 15\%.  Typically, a change of the linewidth at the highest effective temperature of 1~$\mu$Hz will increase $s$ by 1.  The sensitivity of the power law index $s$ to the high effective temperatures  also explains why the index does not vary much when other data sets obtained at lower effective temperature are included (The data sets I and II from the MLE fitters are at low effective temperature).

   \begin{table}[t]
\caption{Parameters of the fit of Eq.~(\ref{Eq1}) and their random errors for linewidth measured at {\it maximum mode height}.}             
\label{table:1}      
\centering                          
\begin{tabular}{c c c c c}        
\hline                 
\hline     
Dataset &$T_{\rm eff}$&$\Gamma_0$ ($\mu$Hz) & $\alpha$ ($\mu$Hz)& s\\    
\hline  
\hline                           
I+II+III+IV+V& Pins.					& 0.35 $\pm$ 0.13 	& 0.91 $\pm$ 0.16 & 13.7 $\pm$ 1.4 \\      
 I+II$\dagger$&	Pins.		&0.32 $\pm$ 0.17		&0.93 $\pm$ 0.20	& 12.7 $\pm$ 2.1 \\
 All &	Pins.			&0.46 $\pm$ 0.09      &0.75 $\pm$ 0.11 & 15.4 $\pm$ 1.3 \\
     \hline     
I+II+III+IV+V& Casa.				& 0.20 $\pm$ 0.14 	& 0.97 $\pm$ 0.17 	& 13.0 $\pm$ 1.4 \\

\hline                                   
\hline  
\end{tabular}
$\dagger${Range for these stars is 5300 K to 6400 K}
\end{table}
   
\begin{table}[!]
\caption{Parameters of the fit of Eq.~(\ref{Eq1}) and their random errors bars for linewidth measured at {\it maximum mode amplitude}.}             
\label{table:2}      
\centering                          
\begin{tabular}{c c c c c}        
\hline                 
\hline     
Dataset &$T_{\rm eff}$&$\Gamma_0$ ($\mu$Hz) & $\alpha$ ($\mu$Hz)& s\\    
\hline  
\hline                           
I+II+III+IV+V& Pins.& 0.64 $\pm$ 0.11 & 0.66 $\pm$ 0.14 & 16.7 $\pm$ 1.8 \\      
 I+II$\dagger$&	Pins.&0.65 $\pm$ 0.10&0.64 $\pm$ 0.13 & 16.1 $\pm$ 2.3 \\
  All&	Pins.			&0.65 $\pm$ 0.09 &0.64 $\pm$ 0.10 & 17.0 $\pm$ 1.4 \\
 \hline     
I+II+III+IV+V& Casa.& 0.49 $\pm$ 0.12 & 0.75 $\pm$ 0.15 & 15.5 $\pm$ 1.6 \\
\hline                                   
\hline  
\end{tabular}
$\dagger${Range for these stars is 5300 K to 6400 K}
\end{table}

\section{Conclusion}
We studied the dependence of linewidth at maximum mode height and amplitude on $T_{\rm eff}$ for two sources effective temperature and for two ways of deriving the linewidth.  We showed using 9 months of {\it Kepler} observations of 42 stars that the mode linewidth at both maximum mode height or maximum amplitude follows a scaling relation based on effective temperature, which is a combination of a power law plus a lower bound.  We stress that this scaling relation is only valid for the cool main-sequence and subgiant stars, and does not have predictive power outside the temperature range of these stars. 


\begin{acknowledgements}
The authors wish to thank the entire {\it Kepler} team, without whom
these results would not be possible. Funding for this Discovery mission is provided by NASA's
Science Mission Directorate.  We also thank all funding
councils and agencies that have supported the activities
of KASC Working Group~1, as well as the International Space Science
Institute (ISSI).  This research was supported in part by the National Science Foundation under Grant No. NSF PHY05-51164.  WJC, GAV and YE acknowledge financial support from the UK Science and Technology Facilities Council (STFC).  RAG and GRD has received funding from the European Community's Seventh Framework Programme (FP7/2007-2013) under grant agreement no. 269194.  MG received financial support from an NSERC Vanier scholarship. This work employed computational facilities provided by ACEnet, the regional high performance computing consortium for universities in Atlantic Canada.  SH acknowledges funding from the Nederlandse Organisatie voor Wetenschappelijk Onderzoek (NWO). GH acknowledges support by the Austrian Science Fund (FWF) project
P21205-N16.  RH acknowledges computing support from the National Solar Observatory.  DS acknowledges the financial support  from the Centre National d'Etudes Spatiales (CNES).
\end{acknowledgements}

\bibliographystyle{aa}
\bibliography{thierrya}

\begin{thebibliography}{37}
\expandafter\ifx\csname natexlab\endcsname\relax\def\natexlab#1{#1}\fi

\bibitem[{{Appourchaux}(2011)}]{Appourchaux2011}
{Appourchaux}, T. 2011, A crash course on data analysis in asteroseismology,
  XXIIth Winter school of the Canary Islands, ArXiv e-prints 1103.5352

\bibitem[{{Baglin}(2006)}]{Baglin2006}
{Baglin}, A. 2006, {The CoRoT mission, pre-launch status, stellar seismology
  and planet finding} (M.Fridlund, A.Baglin, J.Lochard and L.Conroy eds, ESA
  SP-1306, ESA Publication Division, Noordwijk, The Netherlands)

\bibitem[{{Ballot} {et~al.}(2011){Ballot}, {Barban}, \& {van't
  Veer-Menneret}}]{Ballot2011}
{Ballot}, J., {Barban}, C., \& {van't Veer-Menneret}, C. 2011, \aap, {\bf 531},
  A124

\bibitem[{{Balmforth}(1992)}]{B92a}
{Balmforth}, N.~J. 1992, \mnras, {\bf 255}, 603

\bibitem[{{Baudin} {et~al.}(2011{\natexlab{a}}){Baudin}, {Barban}, {Belkacem},
  {Hekker}, {Morel}, {Samadi}, {Benomar}, {Goupil}, {Carrier}, {Ballot},
  {Deheuvels}, {De Ridder}, {Hatzes}, {Kallinger}, \& {Weiss}}]{Baudin2011}
{Baudin}, F., {Barban}, C., {Belkacem}, K., {et~al.} 2011{\natexlab{a}}, \aap,
  {\bf 529}, A84

\bibitem[{{Baudin} {et~al.}(2011{\natexlab{b}}){Baudin}, {Barban}, {Belkacem},
  {Hekker}, {Morel}, {Samadi}, {Benomar}, {Goupil}, {Carrier}, {Ballot},
  {Deheuvels}, {De Ridder}, {Hatzes}, {Kallinger}, \& {Weiss}}]{Baudin2011C}
{Baudin}, F., {Barban}, C., {Belkacem}, K., {et~al.} 2011{\natexlab{b}}, \aap,
  {\bf 535}, C1

\bibitem[{{Bedding}(2011)}]{Bedding2011}
{Bedding}, T.~R. 2011, Solar-like Oscillations: An Observational Perspective,
  XXIIth Winter school of the Canary Islands, ArXiv e-prints 1107.1723

\bibitem[{{Belkacem} {et~al.}(2011){Belkacem}, {Goupil}, {Dupret}, {Samadi},
  {Baudin}, {Noels}, \& {Mosser}}]{Belkacem2011}
{Belkacem}, K., {Goupil}, M.~J., {Dupret}, M.~A., {et~al.} 2011, \aap, {\bf
  530}, A142

\bibitem[{{Benomar} {et~al.}(2009){Benomar}, {Appourchaux}, \&
  {Baudin}}]{Benomar2009}
{Benomar}, O., {Appourchaux}, T., \& {Baudin}, F. 2009, \aap, {\bf 506}, 15

\bibitem[{{Borucki} {et~al.}(2009){Borucki}, {Koch}, {Jenkins}, {Sasselov},
  {Gilliland}, {Batalha}, {Latham}, {Caldwell}, {Basri}, {Brown},
  {Christensen-Dalsgaard}, {Cochran}, {DeVore}, {Dunham}, {Dupree}, {Gautier},
  {Geary}, {Gould}, {Howell}, {Kjeldsen}, {Lissauer}, {Marcy}, {Meibom},
  {Morrison}, \& {Tarter}}]{Borucki2009}
{Borucki}, W.~J., {Koch}, D., {Jenkins}, J., {et~al.} 2009, Science, {\bf 325},
  709

\bibitem[{{Campante} {et~al.}(2011){Campante}, {Handberg}, {Mathur},
  {Appourchaux}, {Bedding}, {Chaplin}, {Garc{\'{\i}}a}, {Mosser}, {Benomar},
  {Bonanno}, {Corsaro}, {Fletcher}, {Gaulme}, {Hekker}, {Karoff}, {R{\'e}gulo},
  {Salabert}, {Verner}, {White}, {Houdek}, {Brand{\~a}o}, {Creevey}, {Do{\v
  g}an}, {Bazot}, {Christensen-Dalsgaard}, {Cunha}, {Elsworth}, {Huber},
  {Kjeldsen}, {Lundkvist}, {Molenda-{\.Z}akowicz}, {Monteiro}, {Stello},
  {Clarke}, {Girouard}, \& {Hall}}]{Campante2011}
{Campante}, T.~L., {Handberg}, R., {Mathur}, S., {et~al.} 2011, \aap, {\bf
  534}, A6

\bibitem[{{Casagrande} {et~al.}(2006){Casagrande}, {Portinari}, \&
  {Flynn}}]{Casa2006}
{Casagrande}, L., {Portinari}, L., \& {Flynn}, C. 2006, \mnras, {\bf 373}, 13

\bibitem[{{Casagrande} {et~al.}(2010){Casagrande}, {Ram{\'{\i}}rez},
  {Mel{\'e}ndez}, {Bessell}, \& {Asplund}}]{Casa2010}
{Casagrande}, L., {Ram{\'{\i}}rez}, I., {Mel{\'e}ndez}, J., {Bessell}, M., \&
  {Asplund}, M. 2010, \aap, {\bf 512}, A54

\bibitem[{{Chaplin} {et~al.}(2000){Chaplin}, {Elsworth}, {Isaak}, {Miller}, \&
  {New}}]{WC2000}
{Chaplin}, W.~J., {Elsworth}, Y., {Isaak}, G.~R., {Miller}, B.~A., \& {New}, R.
  2000, \mnras, {\bf 313}, 32

\bibitem[{{Chaplin} {et~al.}(2009){Chaplin}, {Houdek}, {Karoff}, {Elsworth}, \&
  {New}}]{Chaplin2009}
{Chaplin}, W.~J., {Houdek}, G., {Karoff}, C., {Elsworth}, Y., \& {New}, R.
  2009, \aap, {\bf 500}, L21

\bibitem[{{Chaplin} {et~al.}(2011){Chaplin}, {Kjeldsen},
  {Christensen-Dalsgaard}, {Basu}, {Miglio}, {Appourchaux}, {Bedding},
  {Elsworth}, {Garc{\'{\i}}a}, {Gilliland}, {Girardi}, {Houdek}, {Karoff},
  {Kawaler}, {Metcalfe}, {Molenda-{\.Z}akowicz}, {Monteiro}, {Thompson},
  {Verner}, {Ballot}, {Bonanno}, {Brand{\~a}o}, {Broomhall}, {Bruntt},
  {Campante}, {Corsaro}, {Creevey}, {Do{\u g}an}, {Esch}, {Gai}, {Gaulme},
  {Hale}, {Handberg}, {Hekker}, {Huber}, {Jim{\'e}nez}, {Mathur}, {Mazumdar},
  {Mosser}, {New}, {Pinsonneault}, {Pricopi}, {Quirion}, {R{\'e}gulo},
  {Salabert}, {Serenelli}, {Aguirre}, {Sousa}, {Stello}, {Stevens}, {Suran},
  {Uytterhoeven}, {White}, {Borucki}, {Brown}, {Jenkins}, {Kinemuchi}, {Van
  Cleve}, \& {Klaus}}]{Chaplin2011}
{Chaplin}, W.~J., {Kjeldsen}, H., {Christensen-Dalsgaard}, J., {et~al.} 2011,
  Science, {\bf 332}, 213

\bibitem[{{Deheuvels} {et~al.}(2010){Deheuvels}, {Bruntt}, {Michel}, {Barban},
  {Verner}, {R{\'e}gulo}, {Mosser}, {Mathur}, {Gaulme}, {Garcia}, {Boumier},
  {Appourchaux}, {Samadi}, {Catala}, {Baudin}, {Baglin}, {Auvergne},
  {Roxburgh}, \& {P{\'e}rez Hern{\'a}ndez}}]{Deheuvels2010}
{Deheuvels}, S., {Bruntt}, H., {Michel}, E., {et~al.} 2010, \aap, {\bf 515},
  A87

\bibitem[{{Feroz} {et~al.}(2009){Feroz}, {Hobson}, \& {Bridges}}]{Feroz2009}
{Feroz}, F., {Hobson}, M.~P., \& {Bridges}, M. 2009, \mnras, {\bf 398}, 1601

\bibitem[{{Fr\"ohlich} {et~al.}(1995){Fr\"ohlich}, {Romero}, {Roth}, {Wehrli},
  {Andersen}, {Appourchaux}, {Domingo}, {Telljohann}, {Berthomieu}, {Delache},
  {Provost}, {Toutain}, {Crommelynck}, {Chevalier}, {Fichot}, {D\"appen},
  {Gough}, {Hoeksema}, {Jim\'enez}, {G\'omez}, {Herreros}, {Cort\'es}, {Jones},
  {Pap}, \& {Willson}}]{CFJR95}
{Fr\"ohlich}, C., {Romero}, J., {Roth}, H., {et~al.} 1995, \solphys, {\bf 162},
  101

\bibitem[{{Garc{\'{\i}}a} {et~al.}(2011){Garc{\'{\i}}a}, {Hekker}, {Stello},
  {Guti{\'e}rrez-Soto}, {Handberg}, {Huber}, {Karoff}, {Uytterhoeven},
  {Appourchaux}, {Chaplin}, {Elsworth}, {Mathur}, {Ballot},
  {Christensen-Dalsgaard}, {Gilliland}, {Houdek}, {Jenkins}, {Kjeldsen},
  {McCauliff}, {Metcalfe}, {Middour}, {Molenda-Zakowicz}, {Monteiro}, {Smith},
  \& {Thompson}}]{RAG2011}
{Garc{\'{\i}}a}, R.~A., {Hekker}, S., {Stello}, D., {et~al.} 2011, \mnras, {\bf
  414}, L6

\bibitem[{{Gilliland} {et~al.}(2010){Gilliland}, {Jenkins}, {Borucki},
  {Bryson}, {Caldwell}, {Clarke}, {Dotson}, {Haas}, {Hall}, {Klaus}, {Koch},
  {McCauliff}, {Quintana}, {Twicken}, \& {van Cleve}}]{Gilliland2010}
{Gilliland}, R.~L., {Jenkins}, J.~M., {Borucki}, W.~J., {et~al.} 2010, \apjl,
  {\bf 713}, L160

\bibitem[{{Gizon} \& {Solanki}(2003)}]{Gizon2003}
{Gizon}, L. \& {Solanki}, S.~K. 2003, \apj, {\bf 589}, 1009

\bibitem[{{Gruberbauer} {et~al.}(2009){Gruberbauer}, {Kallinger}, {Weiss}, \&
  {Guenther}}]{Gruberbauer2009}
{Gruberbauer}, M., {Kallinger}, T., {Weiss}, W.~W., \& {Guenther}, D.~B. 2009,
  \aap, {\bf 506}, 1043

\bibitem[{{Handberg} \& {Campante}(2011)}]{Handberg2011}
{Handberg}, R. \& {Campante}, T.~L. 2011, \aap, {\bf 527}, A56

\bibitem[{{Harvey}(1985)}]{Harvey85}
{Harvey}, J. 1985, in Future missions in solar, heliospheric and space plasma
  physics, ESA SP-235, ed. E.Rolfe \& B.Battrick (ESA Publications Division,
  Noordwijk, The Netherlands), 199

\bibitem[{{Houdek} {et~al.}(1999){Houdek}, {Balmforth},
  {Christensen-Dalsgaard}, \& {Gough}}]{GH99}
{Houdek}, G., {Balmforth}, N.~J., {Christensen-Dalsgaard}, J., \& {Gough},
  D.~O. 1999, \aap, {\bf 351}, 582

\bibitem[{{Jenkins} {et~al.}(2010){Jenkins}, {Caldwell}, {Chandrasekaran},
  {Twicken}, {Bryson}, {Quintana}, {Clarke}, {Li}, {Allen}, {Tenenbaum}, {Wu},
  {Klaus}, {Middour}, {Cote}, {McCauliff}, {Girouard}, {Gunter}, {Wohler},
  {Sommers}, {Hall}, {Uddin}, {Wu}, {Bhavsar}, {Van Cleve}, {Pletcher},
  {Dotson}, {Haas}, {Gilliland}, {Koch}, \& {Borucki}}]{Jenkins2010}
{Jenkins}, J.~M., {Caldwell}, D.~A., {Chandrasekaran}, H., {et~al.} 2010,
  \apjl, {\bf 713}, L87

\bibitem[{{Kjeldsen} \& {Bedding}(1995)}]{HK1995}
{Kjeldsen}, H. \& {Bedding}, T.~R. 1995, \aap, {\bf 293}, 87

\bibitem[{{Kjeldsen} \& {Bedding}(2011)}]{KB2011}
{Kjeldsen}, H. \& {Bedding}, T.~R. 2011, \aap, {\bf 529}, L8

\bibitem[{{Marigo} {et~al.}(2008){Marigo}, {Girardi}, {Bressan}, {Groenewegen},
  {Silva}, \& {Granato}}]{Marigo2008}
{Marigo}, P., {Girardi}, L., {Bressan}, A., {et~al.} 2008, \aap, {\bf 482}, 883

\bibitem[{{Mathur} {et~al.}(2011){Mathur}, {Handberg}, {Campante},
  {Garc{\'{\i}}a}, {Appourchaux}, {Bedding}, {Mosser}, {Chaplin}, {Ballot},
  {Benomar}, {Bonanno}, {Corsaro}, {Gaulme}, {Hekker}, {R{\'e}gulo},
  {Salabert}, {Verner}, {White}, {Brand{\~a}o}, {Creevey}, {Do{\u g}an},
  {Elsworth}, {Huber}, {Hale}, {Houdek}, {Karoff}, {Metcalfe},
  {Molenda-{\.Z}akowicz}, {Monteiro}, {Thompson}, {Christensen-Dalsgaard},
  {Gilliland}, {Kawaler}, {Kjeldsen}, {Quintana}, {Sanderfer}, \&
  {Seader}}]{Mathur2011}
{Mathur}, S., {Handberg}, R., {Campante}, T.~L., {et~al.} 2011, \apj, {\bf
  733}, 95

\bibitem[{{Metcalfe} {et~al.}(2010){Metcalfe}, {Monteiro}, {Thompson},
  {Molenda-{\.Z}akowicz}, {Appourchaux}, {Chaplin}, {Do{\u g}an},
  {Eggenberger}, {Bedding}, {Bruntt}, {Creevey}, {Quirion}, {Stello},
  {Bonanno}, {Silva Aguirre}, {Basu}, {Esch}, {Gai}, {Di Mauro}, {Kosovichev},
  {Kitiashvili}, {Su{\'a}rez}, {Moya}, {Piau}, {Garc{\'{\i}}a}, {Marques},
  {Frasca}, {Biazzo}, {Sousa}, {Dreizler}, {Bazot}, {Karoff}, {Frandsen},
  {Wilson}, {Brown}, {Christensen-Dalsgaard}, {Gilliland}, {Kjeldsen},
  {Campante}, {Fletcher}, {Handberg}, {R{\'e}gulo}, {Salabert}, {Schou},
  {Verner}, {Ballot}, {Broomhall}, {Elsworth}, {Hekker}, {Huber}, {Mathur},
  {New}, {Roxburgh}, {Sato}, {White}, {Borucki}, {Koch}, \&
  {Jenkins}}]{Metcalfe2010}
{Metcalfe}, T.~S., {Monteiro}, M.~J.~P.~F.~G., {Thompson}, M.~J., {et~al.}
  2010, \apj, {\bf 723}, 1583

\bibitem[{{Pinsonneault} {et~al.}(2011){Pinsonneault}, {An},
  {Molenda-{\.Z}akowicz}, {Chaplin}, {Metcalfe}, \& {Bruntt}}]{Pins2011}
{Pinsonneault}, M.~H., {An}, D., {Molenda-{\.Z}akowicz}, J., {et~al.} 2011,
  ArXiv e-prints, 1110.4456

\bibitem[{{Samadi}(2009)}]{Samadi2009}
{Samadi}, R. 2009, ArXiv e-prints, 0912.0817

\bibitem[{{Scargle}(1982)}]{Scargle82}
{Scargle}, J.~D. 1982, \apj, {\bf 263}, 835

\bibitem[{{Toutain} \& {Appourchaux}(1994)}]{TTTA94}
{Toutain}, T. \& {Appourchaux}, T. 1994, \aap, {\bf 289}, 649

\bibitem[{{Toutain} {et~al.}(2005){Toutain}, {Elsworth}, \&
  {Chaplin}}]{TT2005b}
{Toutain}, T., {Elsworth}, Y., \& {Chaplin}, W.~J. 2005, \aap, {\bf 433}, 713

\end{thebibliography}

\Online

\begin{table*}[!]
\caption{Natural logarithm of the linewidth measured at maximum mode height with their error bars for each star, together with the frequency of the maximum, the effective temperature of \citet{Pins2011} and of \citet{Casa2010, Casa2006}, with their respective error bars.}             
\label{table:2}      
\centering                          
\begin{tabular}{r c c c c c}        
\hline                 
KIC number &$T_{\rm eff}^{\rm Pins}$&$T_{\rm eff}^{\rm Cas}$&$\nu_{\rm max}$&$\gamma$ ($\ln \mu$Hz) & Fitter\\    
\hline                           
     1435467&        6541 $\pm$          126&        6433 $\pm$           58&      1414.3&      1.422 $\pm$     0.073&IAS\\
     2837475&        6710 $\pm$           61&        6664 $\pm$           92&      1585.3&      2.228 $\pm$     0.072&IAS\\
     3424541&        6460 $\pm$           55&        6723 $\pm$           83&      678.8&      1.480 $\pm$      0.112&IAS\\
     3427720&        5970 $\pm$           52&        6100 $\pm$           80&      2684.6&     0.542 $\pm$     0.093&IAS\\
     3733735&        6720 $\pm$           56&        6827 $\pm$           96&      2026.9&      2.227 $\pm$      0.102&IAS\\
     3735871&        6220 $\pm$           61&        6298 $\pm$           67&      2747.4&      1.012 $\pm$      0.137&IAS\\
     6116048&        6020 $\pm$           51&        6073 $\pm$           69&      2150.0&     0.420 $\pm$     0.072&IAS\\
     6508366&        6480 $\pm$           56&        6379 $\pm$           90&      979.8&      1.599 $\pm$     0.074&IAS\\
     6603624&        5610 $\pm$           51&        5672 $\pm$           58&      2367.0&    -0.423 $\pm$     0.078&IAS\\
     6679371&        6590 $\pm$           56&        6473 $\pm$           89&      854.0&      1.623 $\pm$     0.062&IAS\\
     6933899&        5820 $\pm$           50&        5837 $\pm$           73&      1393.9&     0.239 $\pm$     0.065&IAS\\
     7103006&        6390 $\pm$           56&        6381 $\pm$           84&      1132.8&      1.415 $\pm$     0.080&IAS\\
     7106245&        6020 $\pm$           51&        6041 $\pm$           69&      2382.8&     0.312 $\pm$      0.182&IAS\\
     7206837&        6360 $\pm$           56&        6428 $\pm$           75&      1509.0&      1.472 $\pm$     0.095&IAS\\
     7871531&        5390 $\pm$           47&        5331 $\pm$           42&      3254.7&     0.122 $\pm$      0.146&IAS\\
     8006161&        5300 $\pm$           46&        5399 $\pm$           41&      3518.5&    -0.354 $\pm$     0.085&IAS\\
     8228742&        6080 $\pm$           51&        6235 $\pm$           76&      1126.9&     0.824 $\pm$     0.070&IAS\\
     8379927&        5990 $\pm$           52&        5965 $\pm$           62&      2684.0&     0.815 $\pm$     0.066&IAS\\
     8394589&        6210 $\pm$           52&        6276 $\pm$           75&      2328.6&     0.942 $\pm$     0.085&IAS\\
     8694723&        6310 $\pm$           56&        6401 $\pm$           73&      1435.3&      1.148 $\pm$     0.051&IAS\\
     9025370&        5660 $\pm$           52&        5737 $\pm$           69&      2848.3&    -0.173 $\pm$      0.188&IAS\\
     9098294&        5960 $\pm$           51&        5984 $\pm$           60&      2334.9&     0.481 $\pm$     0.089&IAS\\
     9139151&        6090 $\pm$           52&        6226 $\pm$           78&      2620.2&      1.040 $\pm$     0.084&IAS\\
     9139163&        6370 $\pm$           56&        6510 $\pm$           90&      1704.5&      1.569 $\pm$     0.055&IAS\\
     9206432&        6470 $\pm$           56&        6677 $\pm$          109&      1903.9&      2.129 $\pm$     0.086&IAS\\
     9410862&        6180 $\pm$           51&        6174 $\pm$           65&      2184.9&     0.732 $\pm$      0.137&IAS\\
     9812850&        6380 $\pm$           55&        6382 $\pm$           95&      1264.4&      1.680 $\pm$     0.078&IAS\\
     9955598&        5450 $\pm$           47&        5492 $\pm$           45&      3453.4&    -0.642 $\pm$      0.180&IAS\\
    10018963&        6230 $\pm$           52&        6154 $\pm$           78&      947.2&     0.854 $\pm$     0.052&IAS\\
    10162436&        6320 $\pm$           53&        6253 $\pm$           77&      1008.6&     0.981 $\pm$     0.064&IAS\\
    10355856&        6540 $\pm$           56&        6595 $\pm$           77&      1280.5&      1.754 $\pm$     0.079&IAS\\
    10454113&        6246 $\pm$           58&        6071 $\pm$           74&      2333.2&      1.245 $\pm$     0.066&IAS\\
    10644253&        6020 $\pm$           51&        6122 $\pm$           69&      2993.2&     0.805 $\pm$      0.137&IAS\\
    10909629&        6490 $\pm$           61&        6420 $\pm$           73&      893.1&      1.220 $\pm$      0.101&IAS\\
    10963065&        6280 $\pm$           51&        6177 $\pm$           67&      2195.5&     0.822 $\pm$     0.064&IAS\\
    11081729&        6600 $\pm$           62&        6696 $\pm$           81&      1803.2&      1.887 $\pm$      0.103&IAS\\
    11244118&        5620 $\pm$           51&        5824 $\pm$           62&      1383.7&   -0.081 $\pm$     0.077&IAS\\
    11253226&        6690 $\pm$           56&        6789 $\pm$           99&      1685.8&      2.166 $\pm$     0.056&IAS\\
    11772920&        5420 $\pm$           51&        5440 $\pm$           44&      3394.7&    -0.241 $\pm$      0.174&IAS\\
    12009504&        6230 $\pm$           51&        6337 $\pm$           71&      1870.5&     0.628 $\pm$     0.092&IAS\\
    12258514&        5950 $\pm$           51&        5967 $\pm$           70&      1517.1&     0.515 $\pm$     0.053&IAS\\
    12317678&        6540 $\pm$           55&        6558 $\pm$           86&      1201.9&      1.594 $\pm$     0.058&IAS\\
     6116048&        6020 $\pm$           51&        6073 $\pm$           69&      2150.0&     0.507 $\pm$     0.053&BIR\\
     6603624&        5610 $\pm$           51&        5672 $\pm$           58&      2367.0&    -0.427 $\pm$     0.042&BIR\\
     6933899&        5820 $\pm$           50&        5837 $\pm$           73&      1321.5&     0.278 $\pm$     0.031&BIR\\
     8006161&        5300 $\pm$           46&        5399 $\pm$           41&      3518.4&    -0.312 $\pm$     0.050&BIR\\
     8228742&        6080 $\pm$           51&        6235 $\pm$           76&      1189.0&     0.884 $\pm$     0.042&BIR\\
     8379927&        5990 $\pm$           52&        5965 $\pm$           62&      2684.0&     0.820 $\pm$     0.052&BIR\\
    10018963&        6230 $\pm$           52&        6154 $\pm$           78&      947.2&     0.958 $\pm$     0.033&BIR\\
    10963065&        6280 $\pm$           51&        6177 $\pm$           67&      2092.4&     0.637 $\pm$     0.054&BIR\\
    11244118&        5620 $\pm$           51&        5824 $\pm$           62&      1312.2&    0.028 $\pm$     0.031&BIR\\
    12009504&        6230 $\pm$           51&        6337 $\pm$           71&      1870.5&     0.706 $\pm$     0.069&BIR\\
    12258514&        5950 $\pm$           51&        5967 $\pm$           70&      1517.1&     0.656 $\pm$     0.040&BIR\\
     3735871&        6220 $\pm$           61&        6298 $\pm$           67&      2747.3&     0.792 $\pm$     0.088&SYD\\
     6508366&        6480 $\pm$           56&        6379 $\pm$           90&      980.0&      1.678 $\pm$     0.039&SYD\\
     6679371&        6590 $\pm$           56&        6473 $\pm$           89&      854.0&      1.472 $\pm$     0.032&SYD\\
     7103006&        6390 $\pm$           56&        6381 $\pm$           84&      1133.1&     0.922 $\pm$     0.056&SYD\\
     9206432&        6470 $\pm$           56&        6677 $\pm$          109&      1864.6&      1.911 $\pm$     0.051&SYD\\
     9812850&        6380 $\pm$           55&        6382 $\pm$           95&      1170.3&      1.408 $\pm$     0.050&SYD\\
    11253226&        6690 $\pm$           56&        6789 $\pm$           99&      1608.9&      1.918 $\pm$     0.040&SYD\\
     1435467&        6541 $\pm$          126&        6789 $\pm$           99&      1205.4&      1.318 $\pm$     0.036&MAR\\
     2837475&        6710 $\pm$           61&        6433 $\pm$           58&      1509.2&      1.987 $\pm$     0.039&MAR\\
     3424541&        6460 $\pm$           55&        6664 $\pm$           92&      677.8&      1.688 $\pm$     0.048&MAR\\
     3733735&        6720 $\pm$           56&        6723 $\pm$           83&      1655.4&      1.685 $\pm$     0.049&MAR\\
    10355856&        6540 $\pm$           56&        6595 $\pm$           77&      1280.4&      2.242 $\pm$     0.074&AAU\\
    10909629&        6490 $\pm$           61&        6420 $\pm$           73&      942.5&      1.657 $\pm$      0.118&AAU\\
    11081729&        6600 $\pm$           62&        6696 $\pm$           81&      1892.2&      2.796 $\pm$     0.052&AAU\\
    12317678&        6540 $\pm$           55&        6789 $\pm$           99&      1166.8&      2.229 $\pm$     0.053&AAU\\
\hline  
\end{tabular}
\label{table_mode_height}
\end{table*}

\begin{table*}[!]
\caption{Natural logarithm of the linewidth measured at maximum amplitude with their error bars for each star, together with the frequency of the maximum, the effective temperature of \citet{Pins2011} and of \citet{Casa2010, Casa2006}, with their respective error bars.}             
\label{table:2}      
\centering                          
\begin{tabular}{r c c c c c}        
\hline                 
KIC number &$T_{\rm eff}^{\rm Pins}$&$T_{\rm eff}^{\rm Cas}$&$\nu_{\rm max}$&$\gamma$ ($\ln \mu$Hz) & Fitter\\    
\hline                           
     1435467&        6541 $\pm$          126&        6433 $\pm$           58&      1344.1&      1.462 $\pm$     0.074&IAS\\
     2837475&        6710 $\pm$           61&        6664 $\pm$           92&      1660.2&      2.292 $\pm$     0.065&IAS\\
     3424541&        6460 $\pm$           55&        6723 $\pm$           83&      841.9&      1.971 $\pm$      0.119&IAS\\
     3427720&        5970 $\pm$           52&        6100 $\pm$           80&      2684.6&     0.542 $\pm$     0.093&IAS\\
      3733735&        6720 $\pm$           56&        6827 $\pm$           96&      2119.0&      2.286 $\pm$     0.099&IAS\\
     3735871&        6220 $\pm$           61&        6298 $\pm$           67&      2747.4&      1.012 $\pm$      0.137&IAS\\
     6116048&        6020 $\pm$           51&        6073 $\pm$           69&      2150.0&     0.420 $\pm$     0.072&IAS\\
     6508366&        6480 $\pm$           56&        6379 $\pm$           90&      979.8&      1.599 $\pm$     0.074&IAS\\
     6603624&        5610 $\pm$           51&        5672 $\pm$           58&      2367.0&    -0.423 $\pm$     0.078&IAS\\
     6679371&        6590 $\pm$           56&        6473 $\pm$           89&      1006.6&      1.851 $\pm$     0.061&IAS\\
     6933899&        5820 $\pm$           50&        5837 $\pm$           73&      1393.9&     0.239 $\pm$     0.065&IAS\\
     7103006&        6390 $\pm$           56&        6381 $\pm$           84&      1251.7&      1.708 $\pm$     0.074&IAS\\
     7106245&        6020 $\pm$           51&        6041 $\pm$           69&      2382.8&     0.312 $\pm$      0.182&IAS\\
     7206837&        6360 $\pm$           56&        6428 $\pm$           75&      1745.1&      1.547 $\pm$     0.095&IAS\\
     7871531&        5390 $\pm$           47&        5331 $\pm$           42&      3254.7&     0.122 $\pm$      0.146&IAS\\
     8006161&        5300 $\pm$           46&        5399 $\pm$           41&      3667.8&    -0.110 $\pm$     0.082&IAS\\
     8228742&        6080 $\pm$           51&        6235 $\pm$           76&      1251.5&     0.881 $\pm$     0.066&IAS\\
     8379927&        5990 $\pm$           52&        5965 $\pm$           62&      2804.1&     0.979 $\pm$     0.064&IAS\\
     8394589&        6210 $\pm$           52&        6276 $\pm$           75&      2437.9&      1.178 $\pm$     0.087&IAS\\
     8694723&        6310 $\pm$           56&        6401 $\pm$           73&      1285.9&      1.195 $\pm$     0.054&IAS\\
     9025370&        5660 $\pm$           52&        5737 $\pm$           69&      2981.0&    0.055 $\pm$      0.161&IAS\\
     9098294&        5960 $\pm$           51&        5984 $\pm$           60&      2334.9&     0.481 $\pm$     0.089&IAS\\
     9139151&        6090 $\pm$           52&        6226 $\pm$           78&      2620.2&      1.04 $\pm$     0.084&IAS\\
     9139163&        6370 $\pm$           56&        6510 $\pm$           90&      1624.1&      1.565 $\pm$     0.056&IAS\\
     9206432&        6470 $\pm$           56&        6677 $\pm$          109&      1820.0&      2.076 $\pm$     0.068&IAS\\
     9410862&        6180 $\pm$           51&        6174 $\pm$           65&      2292.1&      1.074 $\pm$      0.145&IAS\\
     9812850&        6380 $\pm$           55&        6382 $\pm$           95&      1264.4&      1.680 $\pm$     0.078&IAS\\
     9955598&        5450 $\pm$           47&        5492 $\pm$           45&      3759.7&     0.207 $\pm$      0.156&IAS\\
    10018963&        6230 $\pm$           52&        6154 $\pm$           78&      947.2&     0.854 $\pm$     0.052&IAS\\
    10162436&        6320 $\pm$           53&        6253 $\pm$           77&      1008.6&     0.981 $\pm$     0.064&IAS\\
    10355856&        6540 $\pm$           56&        6595 $\pm$           77&      1280.5&      1.754 $\pm$     0.079&IAS\\
    10454113&        6246 $\pm$           58&        6071 $\pm$           74&      2333.2&      1.245 $\pm$     0.066&IAS\\
    10644253&        6020 $\pm$           51&        6122 $\pm$           69&      2993.2&     0.805 $\pm$      0.137&IAS\\
    10909629&        6490 $\pm$           61&        6420 $\pm$           73&      844.0&      1.156 $\pm$      0.101&IAS\\
    10963065&        6280 $\pm$           51&        6177 $\pm$           67&      2195.5&     0.822 $\pm$     0.064&IAS\\
    11081729&        6600 $\pm$           62&        6696 $\pm$           81&      1922.7&      1.981 $\pm$     0.097&IAS\\
    11244118&        5620 $\pm$           51&        5824 $\pm$           62&      1383.7&   -0.081 $\pm$     0.077&IAS\\
    11253226&        6690 $\pm$           56&        6789 $\pm$           99&      1685.8&      2.166 $\pm$     0.056&IAS\\
    11772920&        5420 $\pm$           51&        5440 $\pm$           44&      3867.5&     0.441 $\pm$      0.229&IAS\\
    12009504&        6230 $\pm$           51&        6337 $\pm$           71&      1870.5&     0.628 $\pm$     0.092&IAS\\
    12258514&        5950 $\pm$           51&        5967 $\pm$           70&      1517.1&     0.515 $\pm$     0.053&IAS\\
    12317678&        6540 $\pm$           55&        6558 $\pm$           86&      1265.4&      1.700 $\pm$     0.056&IAS\\
     6116048&        6020 $\pm$           51&        6073 $\pm$           69&      2150.0&     0.507 $\pm$     0.053&BIR\\
     6603624&        5610 $\pm$           51&        5672 $\pm$           58&      2367.0&    -0.427 $\pm$     0.042&BIR\\
     6933899&        5820 $\pm$           50&        5837 $\pm$           73&      1321.5&     0.278 $\pm$     0.031&BIR\\
     8006161&        5300 $\pm$           46&        5399 $\pm$           41&      3667.8&    0.020 $\pm$     0.046&BIR\\
     8228742&        6080 $\pm$           51&        6235 $\pm$           76&      1189.0&     0.884 $\pm$     0.042&BIR\\
     8379927&        5990 $\pm$           52&        5965 $\pm$           62&      2804.1&     0.999 $\pm$     0.051&BIR\\
    10018963&        6230 $\pm$           52&        6154 $\pm$           78&      947.2&     0.958 $\pm$     0.033&BIR\\
    10963065&        6280 $\pm$           51&        6177 $\pm$           67&      2195.5&     0.804 $\pm$     0.053&BIR\\
    11244118&        5620 $\pm$           51&        5824 $\pm$           62&      1383.7&    0.060 $\pm$     0.029&BIR\\
    12009504&        6230 $\pm$           51&        6337 $\pm$           71&      1870.5&     0.706 $\pm$     0.069&BIR\\
    12258514&        5950 $\pm$           51&        5967 $\pm$           70&      1517.1&     0.656 $\pm$     0.040&BIR\\
     3735871&        6220 $\pm$           61&        6298 $\pm$           67&      2870.1&     0.812 $\pm$     0.081&SYD\\
     6508366&        6480 $\pm$           56&        6379 $\pm$           90&      980.0&      1.678 $\pm$     0.039&SYD\\
     6679371&        6590 $\pm$           56&        6473 $\pm$           89&      956.3&      1.66 $\pm$     0.031&SYD\\
     7103006&        6390 $\pm$           56&        6381 $\pm$           84&      1133.1&     0.922 $\pm$     0.056&SYD\\
     9206432&        6470 $\pm$           56&        6677 $\pm$          109&      1864.6&      1.911 $\pm$     0.051&SYD\\
     9812850&        6380 $\pm$           55&        6382 $\pm$           95&      1298.6&      1.466 $\pm$     0.047&SYD\\
    11253226&        6690 $\pm$           56&        6789 $\pm$           99&      1685.4&      1.997 $\pm$     0.038&SYD\\
     1435467&        6541 $\pm$          126&        6789 $\pm$           99&      1344.1&      1.508 $\pm$     0.029&MAR\\
     2837475&        6710 $\pm$           61&        6433 $\pm$           58&      1735.4&      2.156 $\pm$     0.035&MAR\\
     3424541&        6460 $\pm$           55&        6664 $\pm$           92&      761.1&      1.854 $\pm$     0.047&MAR\\
     3733735&        6720 $\pm$           56&        6723 $\pm$           83&      2027.2&      2.256 $\pm$     0.043&MAR\\
    10355856&        6540 $\pm$           56&        6595 $\pm$           77&      1346.9&      2.218 $\pm$     0.074&AAU\\
    10909629&        6490 $\pm$           61&        6420 $\pm$           73&      843.9&      1.729 $\pm$      0.117&AAU\\
    11081729&        6600 $\pm$           62&        6696 $\pm$           81&      1982.6&      2.792 $\pm$     0.053&AAU\\
    12317678&        6540 $\pm$           55&        6789 $\pm$           99&      1230.9&      2.179 $\pm$     0.055&AAU\\
\hline  
\end{tabular}
\label{table_amplitude}
\end{table*}

\end{document}